\documentclass[%
reprint,
preprintnumbers,
showpacs,
nofootinbib,
amsmath,amssymb,
aps,
prd,
]{revtex4-1}

\usepackage[dvipdfmx]{graphicx}
\usepackage{dcolumn}
\usepackage{bm}
\usepackage{braket}
\usepackage[colorlinks=true,linkcolor=blue,citecolor=blue,urlcolor=blue]{hyperref}

\newcommand{\ito}[1]{\textcolor{black}{#1}}
\newcommand{\itou}[1]{\textcolor{black}{#1}}

\newcommand{\ikeda}[1]{\textcolor{black}{#1}}

\begin{document}

\preprint{NCTS-HEPAP/2101, KOBE-COSMO-21-03, APS/123-QED}

\title{Axion search with quantum nondemolition detection of magnons}

\author{Tomonori Ikeda}
\email{ikeda.tomonori.2s@kyoto-u.ac.jp}
\affiliation{
Department of Physics, Kyoto University, Kita-Shirakawa, Sakyo-ku, Kyoto 606-8502, Japan
}
\author{Asuka Ito}
\email{asuka.i.aa@m.titech.ac.jp}
\affiliation{
Physics Division, National Center for Theoretical Sciences, Hsinchu, 30013, Taiwan
}
\author{Kentaro Miuchi}
\email{miuchi@phys.sci.kobe-u.ac.jp}
\author{Jiro Soda}
\email{jiro@phys.sci.kobe-u.ac.jp}
\author{Hisaya Kurashige}
\affiliation{
Department of Physics, Kobe University,
Rokkodaicho, Nada-ku, Hyogo 657-8501, Japan
}

\author{Yutaka Shikano}
\email{yshikano@gunma-u.ac.jp}
\affiliation{
Graduate School of Science and Technology, Gunma University, Aramaki, Maebashi, Gunma 371-8510, Japan
 }
\affiliation{
Institute for Quantum Studies, Chapman University, CA 92866, USA
}
\affiliation{JST PRESTO, Honcho, Kawaguchi, Saitama 332-0012, Japan}
\affiliation{
Quantum Computing Center, Keio University, Hiyoshi, Kohoku-ku, Yokohama, Kanagawa 223-8522, Japan
 }
\affiliation{
Research Center for Advanced Science and Technology, The University of Tokyo, Meguro-ku, Tokyo 153-8904, Japan
}

\date{\today}
\begin{abstract}
The axion provides a solution for the strong CP problem and is one of the leading candidates for dark matter. 
This paper proposes an axion detection scheme based on quantum nondemolition detection of magnon, i.e., quanta of collective spin excitations in solid, which is expected to be excited by the axion--electron interaction predicted by the Dine-Fischer-Srednicki-Zhitnitsky~(DFSZ) model. The prototype detector is composed of a ferrimagnetic sphere as an electronic spin target and a superconducting qubit. Both of these are embedded inside a microwave cavity, which leads to a coherent effective interaction between the uniform magnetostatic mode in the ferrimagnetic crystal and the qubit. 
An upper limit for the coupling constant between an axion and an electron is obtained as
\itou{$g_{aee}<2.6\times10^{-6}$}
at the 95\% confidence level for the axion mass of $33.117$~$\mu$eV$<m_{a}<33.130$~$\mu$eV.
\end{abstract}


\maketitle

\section{Introduction}
The Standard Model successfully predicted the existence of Higgs bosons~\cite{20121ATLAS,201230CMS,201388}. However, several long-standing problems in particle physics remain to be explained beyond the Standard Model or its alternative. For example, the theory of quantum chromodynamics requires an extremely fine tuning of parameters to explain the experimentally-observed electric dipole moment of the neutron~\cite{Crewther:1979pi,Afach:2015sja}. This issue is considered to be the strong CP problem. To solve this, Peccei and Quinn proposed a global $U(1)$ symmetry that is broken at a high energy scale $F_{a}$ allowing for the restoration of the CP symmetry, which consequently gives rise to a new pseudoscalar boson called 
axion~\cite{PhysRevLett.38.1440,PhysRevLett.40.223,Wilczek:1977pj}. Axions can provide a significant fraction of the dark matter~(DM)~\cite{PhysRevLett.50.925}. Therefore, axion DM research is an important field for astrophysics and physics beyond the Standard Model.

The axion model is classified into the Kim-Shifman-Vainshtein-Zakharov (KSVZ) 
model~\cite{Kim:1979if,Shifman:1979if}, where axions couple with photons and hadrons, and the Dine-Fischler-Srednicki-Zhitnitsky (DFSZ) model~\cite{Dine:1981rt,Zhitnitsky:1980tq}, where axions also couple with electrons~\cite{DINE1981199}. 
Most experiments for the axion DM research conducted so far were based on the axion--photon coupling through the Primakoff effect~\cite{PhysRevLett.51.1415}. Using this approach, the ADMX experiment excluded the axion mass range of $1.9\ \mu{\rm eV}<m_{a}<3.7\ \mu{\rm eV}$ for the KVSZ model and $2.66\ \mu{\rm eV}<m_{a}<2.81\ \mu{\rm eV}$ for the DFSZ model~\cite{PhysRevLett.120.151301}. 
Alternatively, detection of the axion--electron coupling can provide strong evidence for the DFSZ model. 
Experiments probing coherent scattering of axions by electrons in the axion mass range of $0.1\ {\rm eV}<m_{a}<1\ {\rm keV}$ were performed~\cite{PhysRevLett.118.261301,201346,PhysRevLett.119.181806}.
This mass range was outside the range of $1\ \mu{\rm eV} < m_{a} < 1\ {\rm meV}$ favored by the cosmological and astrophysical bounds~\cite{TURNER199067,RAFFELT19901}. 
An instrument sensitive to the axion--electron coupling in the frequency range of $1\ {\rm GHz} < f_{a} < 1\ {\rm THz}$ is required to probe the axion mass in the favorable range, $1\ \mu{\rm eV}<m_{a}<1\ {\rm meV}$ as the axion mass and frequency are related by
\begin{align}
f_{a}=\frac{\omega_{a}}{2\pi}=\frac{m_{a}c^{2}}{h} \simeq 0.24\left(\frac{m_{a}}{1.0\ \rm{\mu eV}}\right)\ \rm{GHz}.
\label{frefre}
\end{align}

A possible route that enables probing the axion--electron coupling is based on the detection of quanta of collective spin excitations, called magnons, in a ferrimagnetic crystal~\cite{BARBIERI1989357}. 
The magnons are to be excited through the axion--electron interaction  under the strong internal magnetic field inside the crystal. 
The first experiments based on this axion--electron coupling utilized the hybridization of the uniformly precessing magnetostatic mode, i.e., Kittel mode, in spherical ferrimagnetic crystals and a microwave cavity mode~\cite{Barbieri:2016vwg,QUAX, Tober, crescini2020axion}. 
In the experiments they detected an emitted microwave field generated from the magnons that were potentially excited through the axion--electron coupling using a linear receiver~\cite{QUAX, Tober, crescini2020axion}. 
An alternative detection method is the magnon counting based on Quantum Non-Demolition (QND) measurements where each measurement does not affect the quantum state of magnons. While linear receivers are ultimately subject to quantum fluctuations, a magnon counting detector is sensitive to the magnon number. Then, the signal-to-noise ratio of the magnon counting detector is only limited by the shot noise on the detected thermal photons, according to Poisson statistics. Especially in Ref.~\cite{PhysRevD.88.035020}, for the case of the axion--photon conversion, it is shown that the photon counting detector can have lower noise than the linear detector under the temperature of 100~mK and the axion and cavity quality factor ratio of $Q_{a}/Q_{c}=20$.
Resolving the magnon number was successfully demonstrated by reaching the strong dispersive coupling between a superconducting qubit and the Kittel mode in a spherical ferrimagnetic crystal~\cite{Lachance-Quirione1603150}. In this paper, the data obtained in the setup of Ref.~\cite{Lachance-Quirione1603150} is reanalyzed. In addition, it reports experimental results for axion DM research using a QND detection technique for an axion mass range of $33.117$~$\mu$eV$<m_{a} <33.130$~$\mu$eV.

\section{Axion detection scheme}
In this section, the theory of axion--electron interaction is first introduced. The axion-induced effective magnetic field generated by the movement of the Earth through axion DM is discussed. After introducing collective spin excitations in a ferrimagnetic crystal, the QND detection scheme of magnons for the axion DM search is described. This is achieved by measuring the absorption spectrum of a superconducting qubit dispersively coupled to the Kittel mode in the ferrimagnetic crystal.

\subsection{Axion-electron interaction}

The axion emerges as a Nambu-Goldstone boson of the broken Peccei-Quinn symmetry~\cite{PhysRevLett.40.223,Wilczek:1977pj}.
In the DFSZ model~\cite{DINE1981199}, the axion field $a(x)$ can interact with an electron field $\psi(x)$ as
\begin{align}
\mathcal{L}_{\rm{int}}=-i g_{aee}a(x)\bar{\psi}(x)\gamma_{5}\psi(x),
\label{inte}
\end{align}
where $g_{aee}$ is a dimensionless coupling constant, inversely proportional to the energy scale of the Peccei-Quinn symmetry breaking. 
Note that we can rewrite the interaction of Eq.~\eqref{inte} as $\tilde{g}_{aee}(\partial_{\mu} a)\bar{\psi}\gamma^{\mu}\gamma_{5}\psi(x)$ using the background Dirac equation, where $\tilde{g}_{aee}=g_{aee}/2m_{e}$. This clearly shows the shift symmetry of the axion field.
In the non-relativistic limit, the interaction term reads
\begin{align}
\mathcal{H}_{\rm{int}}\simeq-\frac{g_{aee} \hbar}{2m_{e}}\hat{\bm{\sigma}}\cdot\bm{\nabla}a=-2\mu_{{\rm B}}\hat{\bm{S}}\cdot     \left(\frac{g_{aee}}{e}\bm{\nabla}a\right),
\label{eq:electron_axion}
\end{align}
where $m_{e}$ is the electron mass, $e$ is the elementary electric charge, $\mu_{{\rm B}} = e\hbar / 2m_{e}$ is the Bohr magneton.
The electron spin operator $\hat{\bm{S}}$ is related to the Pauli 
matrices~$\hat{\bm{\sigma}}$ with $\hat{\bm{S}} = \hat{\bm{\sigma}} /2 $.
The term in the parentheses can be considered as an effective magnetic field 
\begin{align}
\bm{B}_{a}=\frac{g_{aee}}{e}\bm{\nabla}a,
\label{mag}
\end{align}
by analogy to the usual interaction between the spin and magnetic fields.
If the DM is composed of axions, this effective magnetic field is ubiquitous around us. 
Importantly, the axion DM is oscillating in time at the frequency $f_{a}$ that is related to its mass according to Eq.~\eqref{frefre}.

\subsection{Axion-induced effective magnetic field}
Since the occupation number of the QCD axion DM is high, it would be classical fields.
As the solution of the Klein-Gordon equation, the axion DM oscillates coherently with the
frequency given by Eq.\,(\ref{frefre}).
The coherence length is determined by the de Broglie wavelength of the axion fields,
$1/m_a v_{{\rm tot}}$, where $v_{{\rm tot}}$ is the relative velocity of the 
axion DM to the Earth (see the appendix \ref{apn}).
Then the coherence time, during which the axion-induced effective magnetic field is coherent, 
can be estimated as $1/m_a v^2_{{\rm tot}}$.
\itou{Notably, the coherence time
is about $10^{-4}$\,s and much longer than the decoherence time of our experiment $\sim 10^{-6}$\,s.}

Let us now estimate the magnitude of the axion-induced effective magnetic field (\ref{mag}).
The spatial gradient of the axion DM is evaluated as 
$\partial_{i}a \simeq m_{a}va$~\cite{Aoki:2016kwl} 
because the proper time of the moving axion DM depends 
on the spatial coordinates of the laboratory frame through the Lorentz transformation.
The amplitude of the effective magnetic field $B_a$ can be estimated from Eq.~\eqref{mag} and the relation $\rho_{\rm{DM}}\sim m_{a}^{2}a^{2}/2$,
\begin{align}
B_{a}\simeq 4.4\times10^{-8}  g_{aee}\left(\frac{\rho_{{\rm DM}}}{0.45\ \rm{GeV/cm^{3}}}\right)^{1/2}\left(\frac{v_{{\rm tot}}}{300\ \rm{km/s}}\right)\ \rm{T},
\label{axmag}
\end{align}
where $\rho_{{\rm DM}}$ is the local DM density.
We see that the amplitude is small because $g_{aee}$, 
which is proportional to the inverse of the energy scale of the Peccei-Quinn symmetry breaking, 
is tiny.
However, as we will see in the next subsection, the coupling constant can be larger effectively by considering a collective spin excitation modes in ferrimagnetic crystals.
\subsection{Collective spin excitations}
Let us consider a ferrimagnetic crystal containing $N$ electron spins. This system is described by the Heisenberg model~\cite{Heisenberg1926}
\begin{align}
\hat{\mathcal{H}}_{{\rm m}-a}=2\mu_{{\rm B}}\sum_{i}\hat{\bm{S}}_{i}\cdot\left(\bm{B}_{0}+\bm{B}_{a}\right)-\sum_{i,j}J_{ij}\hat{\bm{S}}_{i} \cdot\hat{\bm{S}}_{j},
\label{hei}
\end{align}
where, $\bm{B}_{0}$ is the external magnetic field and $i$ labels each spins. 
The second term represents the exchange interaction between the neighboring spins with the strength $J_{ij}$. 
Considering an external magnetic field $\bm{B}_{0}$ along the $z$-axis and assuming, without loss of generality, that the direction of the effective magnetic field \ito{lies in the $z$-$x$ plane, we write
\begin{align}
\bm{B}_{0}=(0,0,B_{0}),\quad \bm{B}_a\simeq( |\bm{B}_a|  \sin\theta,0,0),
\label{magg}
\end{align}
where we neglected the $z$-component of $\bm{B}_{a}$ because it is much smaller than $B_{0}$.}
Here, $\theta$ is the angle between the external and effective magnetic fields.

The effective magnetic field is considered to be uniform throughout the sample. 
Thus, the effective magnetic field can be written as
\begin{align}
\bm{B}_{a}(t)=\frac{B_{a}\sin\theta}{2}\left(e^{-i\omega_{a}t}+e^{i\omega_{a}t}\right)\left(1,0,0\right).
\label{ex}
\end{align}
Substituting Eqs.~\eqref{magg} and \eqref{ex} into Eq.~\eqref{hei} yields
\begin{align}
\hat{\mathcal{H}}_{{\rm m}-a}&=2\mu_{{\rm B}}\sum_{i}\left[\hat{S}_{i}^{z}B_{0}+\frac{B_{a}\sin\theta}{4}\left(\hat{S}_{i}^{-}e^{-i\omega_{a}t}+\hat{S}_{i}^{+}e^{i\omega_{a}t}\right)\right]\nonumber\\
&-\sum_{i,j}J_{ij}\hat{\bm{S}}_{i}\cdot\hat{\bm{S}}_{j},
\label{spi}
\end{align}
where $S_{j}^{\pm}=S_{j}^{x}\pm i S_{j}^{y}$ are the spin ladder operators. 
The second term in the right-hand side of Eq.~\eqref{spi} shows that the axion DM excites the spins if the frequency of the effective magnetic field~$\omega_{a}$ is equal to the Larmor frequency $\omega_\mathrm{m}\equiv 2\mu_{{\rm B}}B_{0}/\hbar$.

The spin system of Eq.~\eqref{spi}, including the axion-induced effective magnetic field, can be rewritten in terms of the bosonic operators $\hat{C}_{i}$ and $\hat{C}^{\dagger}_{i}$, which satisfies the commutation relation $[\hat{C}_{i},\hat{C}^{\dagger}_{j}]=\delta_{ij}$, using the Holstein-Primakoff transformation~\cite{Holstein:1940zp}:
\begin{align}
\hat{S}_{i}^{z}&=\frac{1}{2}-\hat{C}_{i}^{\dagger}\hat{C}_{i},\nonumber\\
\label{pri}
\hat{S}_{i}^{+}&=\sqrt{1- \hat{C}_{i}^{\dagger}\hat{C}_{i}}\ \hat{C}_{i},\\
\hat{S}_{i}^{-}&=\hat{C}_{i}^{\dagger}\sqrt{1-\hat{C}_{i}^{\dagger}\hat{C}_{i}}\nonumber.
\end{align}
This introduces spin waves with a dispersion relation determined by the amplitude of the external magnetic field~$B_{0}$ and the amplitudes $J_{ij}$ of the ferrimagnetic exchange interaction. 
Magnons are quanta of the spin-wave modes. Furthermore, provided that the contributions from the surface of the sample are negligible, one can expand the bosonic operators in terms of plane waves as follows:
\begin{align}
\hat{C}_{i}=\frac{1}{\sqrt{N}}\sum_{\bm{k}}e^{-i\bm{k}\cdot\bm{r}_i}\hat{c}_{\bm{k}}.
\label{bbb}
\end{align}
Here, $\bm{r}_{i}$ is the position vector of spin $i$, and $\hat{c}_{\bm{k}}$ annihilates a magnon from the mode with a wave vector $\bm{k}$. 
As the effective magnetic field is induced by axions, it is supposed to be homogeneous over the ferrimagnetic crystal for a typical experiment. 
Then, only the uniform magnetostatic mode, Kittel mode, can be excited as long as the contributions from the surface are negligible~\cite{PhysRev.110.1295}. 
Substituting Eqs.~\eqref{pri} and \eqref{bbb} into the Hamiltonian of Eq.~\eqref{spi} yields
\begin{align}
\label{dri}
\hat{\mathcal{H}}_{{\rm m}-a}&\equiv\hat{\mathcal{H}}_{{\rm m}}+\hat{\mathcal{H}}_{a},\\
\hat{\mathcal{H}}_{{\rm m}}&=\hbar\omega_{\rm{m}}\hat{c}^{\dagger}\hat{c},\\
\hat{\mathcal{H}}_{a}&=g\mu_{{\rm B}}\frac{B_{a}\sin\theta}{4}\sqrt{N}\left(\hat{c}^{\dagger}e^{-i\omega_{a}t}+\hat{c}e^{i\omega_{a}t}\right),
\end{align}
where $\hat{c}\equiv\hat{c}_{k=0}$ and $\overline{n}_\mathrm{m}\equiv\langle\hat{c}^{\dagger}\hat{c}\rangle\ll N$ are assumed. One can see that the coupling strength \ito{has a huge factor of $\sqrt{N}$ due to 
collective spin dynamics}. This effect enables one to potentially detect the small effective magnetic field $B_{a}$ oscillating at $\omega_a$. 

\subsection{Quantum nondemolition detection of magnons}

\begin{figure}[]
\centering
\includegraphics[width=8.5cm]{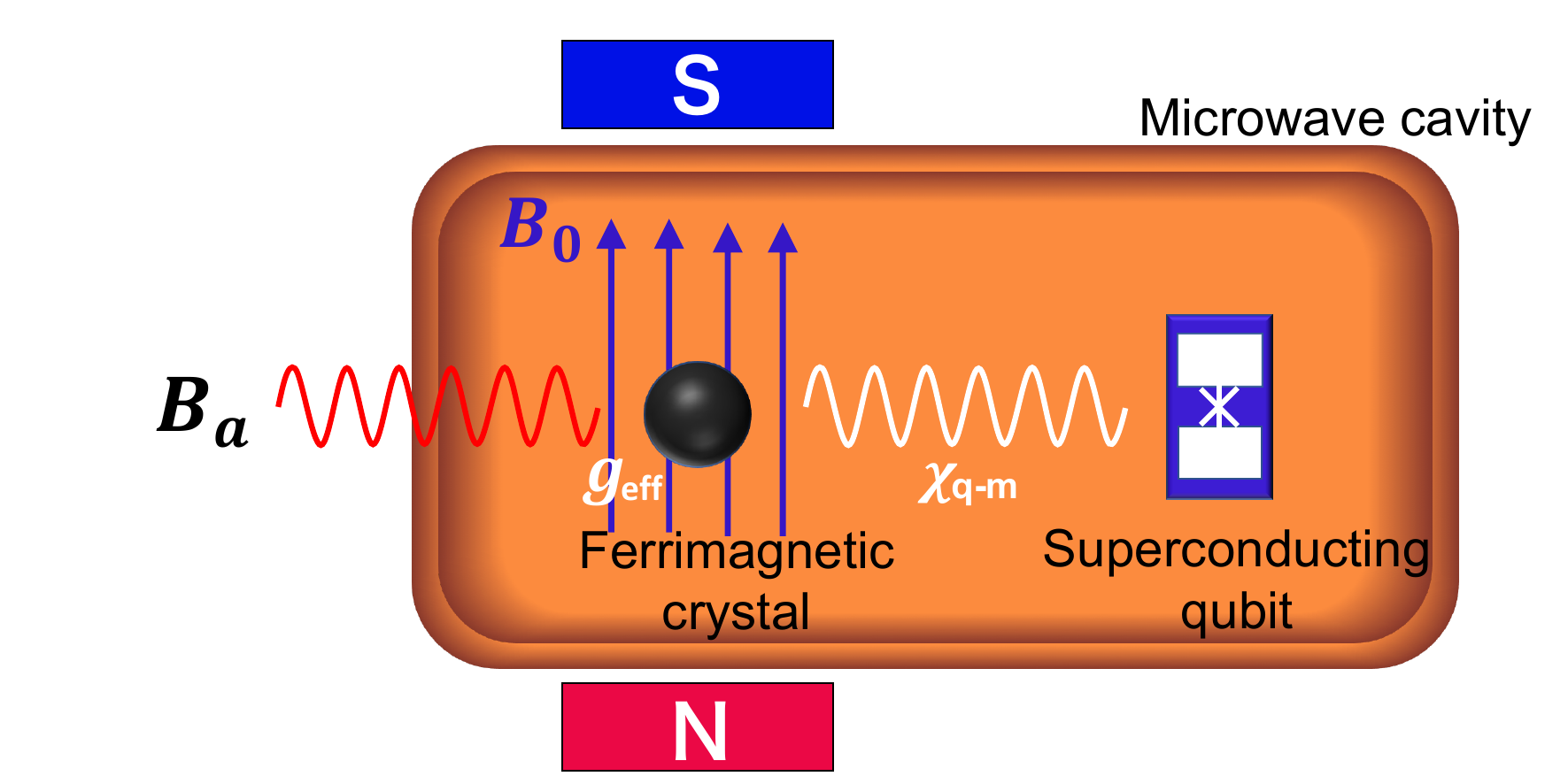}
\caption{Schematic illustration of the detector. A spherical ferrimagnetic crystal and a transmon-type superconducting qubit are coherently coupled through a microwave cavity. 
The effective magnetic field of the axion DM coherently drives the uniform spin-precession mode (Kittel mode) in the ferrimagnetic crystal with an effective coupling constant $g_{\rm{eff}}$. 
In the strong dispersive regime, each magnon excited in the Kittel mode shifts the resonance frequency of the qubit by $2\chi_{{\rm q}-{\rm m}}+\Delta_{a}$, where $\chi_{{\rm q}-{\rm m}}$ is the dispersive shift and $\Delta_{a}=\omega_{{\rm m}}^{g}-\omega_{a}$ is the detuning between the frequency $\omega_a$ of the axion-induced effective magnetic field and the frequency $\omega_{\rm m}^{g}$ of the Kittel mode with the qubit in the ground state $|g\rangle$.}
\label{fig:scheme}
\end{figure}

To detect the excitation of a magnon in the Kittel mode, a hybrid system schematically shown in Fig.~\ref{fig:scheme} is used.
This system consists of a spherical ferrimagnetic crystal and a superconducting qubit~\cite{conducting_qubit}. 
They are individually coupled to the modes of a microwave cavity
through magnetic and electric dipole interactions, respectively.
This leads to an effective coherent coupling between the Kittel mode and the qubit~\cite{TABUCHI2016729,Tabuchi405,Lachance-Quirione1603150,Lachance-Quirion2019, lachancequirion2019entanglementbased}. The transmon-type superconducting qubit~\cite{PhysRevA.76.042319} can be described as an anharmonic oscillator through the Hamiltonian
\begin{align}
\hat{\mathcal{H}}_{\rm q}/\hbar=\left(\omega_{\rm q}-\frac{\alpha}{2}\right)\hat{\mathit{q}}^{\dagger}\hat{\mathit{q}}+\frac{\alpha}{2}\left(\hat{\mathit{q}}^{\dagger}\hat{\mathit{q}}\right)^2,
\label{qubit}
\end{align}
where $\omega_{\rm q}$ is the frequency of the transition between the ground and first-excited states of the qubit, $\ket{g}$ and $\ket{e}$, respectively. The creation and annihilation operators for the qubit are, respectively, $\hat{\mathit{q}}^{\dagger}$ and $\hat{\mathit{q}}$. 
Furthermore, the anharmonicity $\alpha<0$ of the qubit is defined such that the frequency of the transition between the first and second excited states is given by $\omega_{\rm q}+\alpha$~\cite{PhysRevA.74.042318}. 
After adiabatically eliminating the microwave cavity modes from the total Hamiltonian of the hybrid system, the effective interaction Hamiltonian between the Kittel mode and the qubit is given by
\begin{align}
\hat{\mathcal{H}}_{{\rm q}-{\rm m}}/\hbar=g_{{\rm q}-{\rm m}}(\hat{\mathit{q}}^\dagger\hat{\mathit{c}}+\hat{\mathit{q}}\hat{\mathit{c}}^{\dagger}),
\label{qm}
\end{align}
where $g_{{\rm q}-{\rm m}}$ is the coupling strength between the Kittel mode and the qubit~\cite{TABUCHI2016729,Tabuchi405,Lachance-Quirione1603150,Lachance-Quirion2019}.

Combining Eqs.~\eqref{dri}, \eqref{qubit} and \eqref{qm}, the Hamiltonian of the hybrid quantum system, including the effective axion-induced effective magnetic field, is given by
\begin{align}
\hat{\mathcal{H}}_{\rm tot}/\hbar&=\omega_{{\rm m}}\hat{\mathit{c}}^{\dagger}\hat{\mathit{c}}+\left(\omega_{\rm q}-\frac{\alpha}{2}\right)\hat{\mathit{q}}^{\dagger}\hat{\mathit{q}}+\frac{\alpha}{2}\left(\hat{\mathit{q}}^{\dagger}\hat{\mathit{q}}\right)^2\nonumber\\
&+g_{{\rm q}-{\rm m}}(\hat{\mathit{q}}^\dagger\hat{\mathit{c}}+\hat{\mathit{q}}\hat{\mathit{c}}^{\dagger})+g_{\rm eff}\left(\hat{c}^{\dagger}e^{-i\omega_{a}t}+\hat{c}e^{i\omega_{a}t}\right).
\label{tot}
\end{align}
Here,
\begin{align}
\hbar g_{{\rm eff}}=2\mu_{{\rm B}}\frac{B_{a}\sin\theta}{4}\sqrt{N},
\label{eq:geff_Ba}
\end{align}
is the effective coupling constant between axions and magnons, which corresponds to the strength of the coherent magnon drive.

Let us consider the dispersive regime corresponding to a detuning $\Delta_{{\rm q}-{\rm m}}\equiv\omega_{{\rm q}}-\omega_{{\rm m}}$ between the qubit frequency~$\omega_{\rm q}$ and the frequency of the Kittel mode $\omega_{\rm m}$. This is much larger than the coupling strength $g_{{\rm q}-{\rm m}}$ such that the exchange of energy between the two systems is highly suppressed~\cite{Lachance-Quirione1603150}. For this limit, the total Hamiltonian of Eq.~\eqref{tot} can be rewritten as
\begin{align}
\hat{\mathcal{H}}_{\rm tot}'/\hbar&\simeq\omega_{{\rm m}}\hat{\mathit{c}}^{\dagger}\hat{\mathit{c}}+\frac{1}{2}\tilde\omega_{\rm q}\hat{\mathit{\sigma}}_{z}+\chi_{{\rm q}-{\rm m}}\hat{\mathit{c}}^{\dagger}\hat{\mathit{c}}\hat{\mathit{\sigma}}_{z}\nonumber\\
&+g_{{\rm eff}}\left(\hat{c}^{\dagger}e^{-i\omega_{a}t}+\hat{c} e^{i\omega_{a}t}\right),
\label{uni}
\end{align}
where $\tilde\omega_{\rm q}=\omega_{\rm q}+\chi_{{\rm q}-{\rm m}}$ is the qubit frequency shifted by the qubit--magnon dispersive shift $\chi_{{\rm q}-{\rm m}}$, which is described by~\cite{PhysRevA.74.042318}.
\begin{align}
\chi_{{\rm q}-{\rm m}}\simeq\frac{\alpha g_{{\rm q}-{\rm m}}^2}{\Delta_{{\rm q}-{\rm m}}\left(\Delta_{{\rm q}-{\rm m}}+\alpha\right)}.\
\end{align}
The qubit--magnon dispersive shift can also be estimated numerically by diagonalizing the Hamiltonian of the hybrid system~\cite{Lachance-Quirione1603150}. The Hamiltonian of Eq.~\eqref{uni} considers only the first two states of the qubit through the Pauli matrices: $\hat{\sigma}_z=\ket{e}\bra{e}-\ket{g}\bra{g}$, $\hat{\sigma}_+=\ket{e}\bra{g}$, and $\hat{\sigma}_-=\ket{g}\bra{e}$. Furthermore, higher order terms in $\left(g_{{\rm q}-{\rm m}}/\Delta_{{\rm q}-{\rm m}}\right)$ are neglected. 
The third term on the right-hand side of  Eq.~\eqref{uni} shows that the qubit frequency depends on the magnon occupancy through an interaction term, which commutes with the Hamiltonian of the Kittel mode. More specifically, the qubit frequency $\tilde\omega_{\rm q}$ shifts by $2\chi_{{\rm q}-{\rm m}}$ for every magnon in the Kittel mode. Therefore, measuring the qubit frequency enables one to perform a QND detection for the magnon number.

\subsection{Qubit spectrum}
The qubit frequency can be determined, for example, by measuring its absorption spectrum
\begin{align}
S(\omega_{\rm s})={\rm Re}\left[\frac{1}{\sqrt{2\pi}}\int^{\infty}_{0}{\rm d}t\ \langle\hat{\sigma}_{-}(t) \hat{\sigma}_{+}(0)\rangle e^{i \omega_{\rm s}t}\right],
\end{align}
where $\omega_{\rm s}$ is the spectroscopy frequency~\cite{PhysRevA.74.042318}. In this subsection, an analytical model for the qubit spectrum in the presence of a dispersive interaction with the Kittel mode of a ferrimagnetic crystal is provided~\cite{PhysRevA.74.042318,Lachance-Quirione1603150}. For this purpose, the Hamiltonian of Eq.~\eqref{uni} is transformed such that the qubit is in a reference frame that rotates at the spectroscopy frequency $\omega_{\rm s}$. Meanwhile, the Kittel mode is in a reference frame that rotates at the axion frequency $\omega_{a}$. 
The Hamiltonian of Eq.~\eqref{uni} becomes
\begin{align}
\hat{\mathcal{H}}/\hbar&=\left(\Delta_{a}+\chi_{{\rm q}-{\rm m}}\right)\hat{\mathit{c}}^{\dagger}\hat{\mathit{c}}+\frac{1}{2}\Delta_{{\rm s}}\hat{\mathit{\sigma}}_{z}+\chi_{{\rm q}-{\rm m}}\hat{\mathit{c}}^{\dagger}\hat{\mathit{c}}\hat{\mathit{\sigma}}_{z}\nonumber\\
&+g_{{\rm eff}}\left(\hat{c}^{\dagger}+\hat{c}\right)+\Omega_{\rm s}\left(\hat\sigma_++\hat\sigma_-\right),
\label{tiltot}
\end{align}
where $\Delta_{a}=\omega_{{\rm m}}^{g}-\omega_{a}$ is the detuning between the frequency $\omega_{{\rm m}}^{g}$ of the Kittel mode with the qubit in the ground state $\ket{g}$ and the axion frequency $\omega_{a}$. 
In addition, $\Delta_{\rm s}=\tilde\omega_{{\rm q}}-\omega_{s}$ is the detuning between the frequency~$\tilde\omega_{{\rm q}}$ of the qubit and the spectroscopy frequency $\omega_{\rm s}$. 
The amplitudes of the driving terms are given by the effective coupling constant $g_{\rm eff}$ and the Rabi frequency $\Omega_{\rm s}$, respectively.
\ito{Furthermore we have to take into account noises in the system.
Then the Hamiltonian should be $\hat{\mathcal{H}} + \hat{\mathcal{H}}_{\rm{noise}}$,
where $\hat{\mathcal{H}}_{\rm{noise}}$ represents the dephasing mechanisms of the system 
due to free electromagnetic fields (reservoir) in the cavity 
and the environment of the qubit \cite{walls,PhysRevA.74.042318}.
Given the Hamiltonian, we can solve the evolution equation for the reduced density matrix 
with an appropriate initial condition and obtain the qubit spectrum as~\cite{PhysRevA.74.042318}}
\begin{align}
S(\omega_{\rm s})=\frac{1}{\pi}\sum_{n_{\rm{m}}=0}^{\infty}\frac{1}{n_{\rm{m}}!}{\rm Re}\left[\frac{(-A)^{n_{\rm{m}}}e^{A}}{\gamma_{\rm{q}}^{(n_{\rm{m}})}/2-i\left(\omega_{\rm s}-\tilde{\omega}_{\rm{q}}^{(n_{\rm{m}})}\right)}\right],
\label{eq:Gambetta}
\end{align}
with
\begin{align}
\omega_{\rm q}^{(n_{\rm m})}&=\tilde{\omega}_{\rm q}+B+2\chi_{{\rm q}-{\rm m}}n_{\rm{m}}, \\
\tilde{\omega}_{\rm{q}}^{(n_{\rm{m}})}&=\omega_{\rm q}^{(n_{\rm m})}+n_{\rm m}\Delta_{a},\\
\gamma_{\rm{q}}^{(n_{\rm{m}})}&=\gamma_{{\rm q}}+\gamma_{{\rm m}}\left(n_{\rm{m}}+D^{\rm ss}\right),\\
A&=D^{\rm ss}\left(\frac{\gamma_{{\rm m}}/2-i(\Delta_{a}+2\chi_{{\rm q}-{\rm m}})}{\gamma_{{\rm m}}/2+i(\Delta_{a}+2\chi_{{\rm q}-{\rm m}})}\right),\\
B&=\chi_{{\rm q}-{\rm m}}\left(\overline{n}^{g}_{{\rm m}}+\overline{n}^{e}_{{\rm m}}-D^{\rm ss}\right),\\
D^{\rm ss}&=\frac{2\left(\overline{n}^{g}_{{\rm m}}+\overline{n}^{e}_{{\rm m}}\right)\chi_{{\rm q}-{\rm m}}^2}{\left(\gamma_{{\rm m}}/2\right)^2+\chi_{{\rm q}-{\rm m}}^2+\left(\chi_{{\rm q}-{\rm m}}+\Delta_{a}\right)^2},\\
\overline{n}^{g}_{{\rm m}}&=\frac{g_{{\rm eff}}^{2}}{(\gamma_{{\rm m}}/2)^{2}+\Delta_{a}^{2}},\label{eq:geff_and_n}\\
\overline{n}^{e}_{{\rm m}}&=\frac{g_{{\rm eff}}^{2}}{(\gamma_{{\rm m}}/2)^{2}+(\Delta_{a}+2\chi_{{\rm q}-{\rm m}})^{2}}.
\end{align}
\ito{Here, $\gamma_{\rm m}$ and $\gamma_{\rm q}$ are linewidth of the Kittel mode and the qubit, respectively.}
Also, $\tilde{\omega}_{\rm{q}}^{(n_{\rm{m}})}$ and $\gamma_{\rm{q}}^{(n_{\rm{m}})}$ are, respectively, the frequency and the linewidth of the qubit with the Kittel mode in the number state $\ket{n_{\rm m}}$ for a given effective coupling constant~$g_{\rm eff}$. 
Equation~(30) [(31)] expresses the steady-state magnon occupancy when the qubit is in the ground (excited) state, which is given by $\overline{n}^{g(e)}_{{\rm m}}$.

In the strong dispersive regime, which corresponds to $2\chi_{{\rm q}-{\rm m}}\gg\gamma_{{\rm m}}$, the qubit spectrum is given by a sum of Lorentzian functions centered at the shifted qubit frequency $\tilde{\omega}_{\rm{q}}^{(n_{\rm{m}})}$. The spectrum explicitly depends on the population of the magnons potentially excited by the axion-induced effective magnetic field $\bm{B}_{a}$. 
Therefore, the 
axion DM can be probed by measuring the spectrum of the qubit that is coupled to the Kittel mode in the ferrimagnetic crystal.

\begin{figure}[t]
\centering
\includegraphics[width=8.5cm]{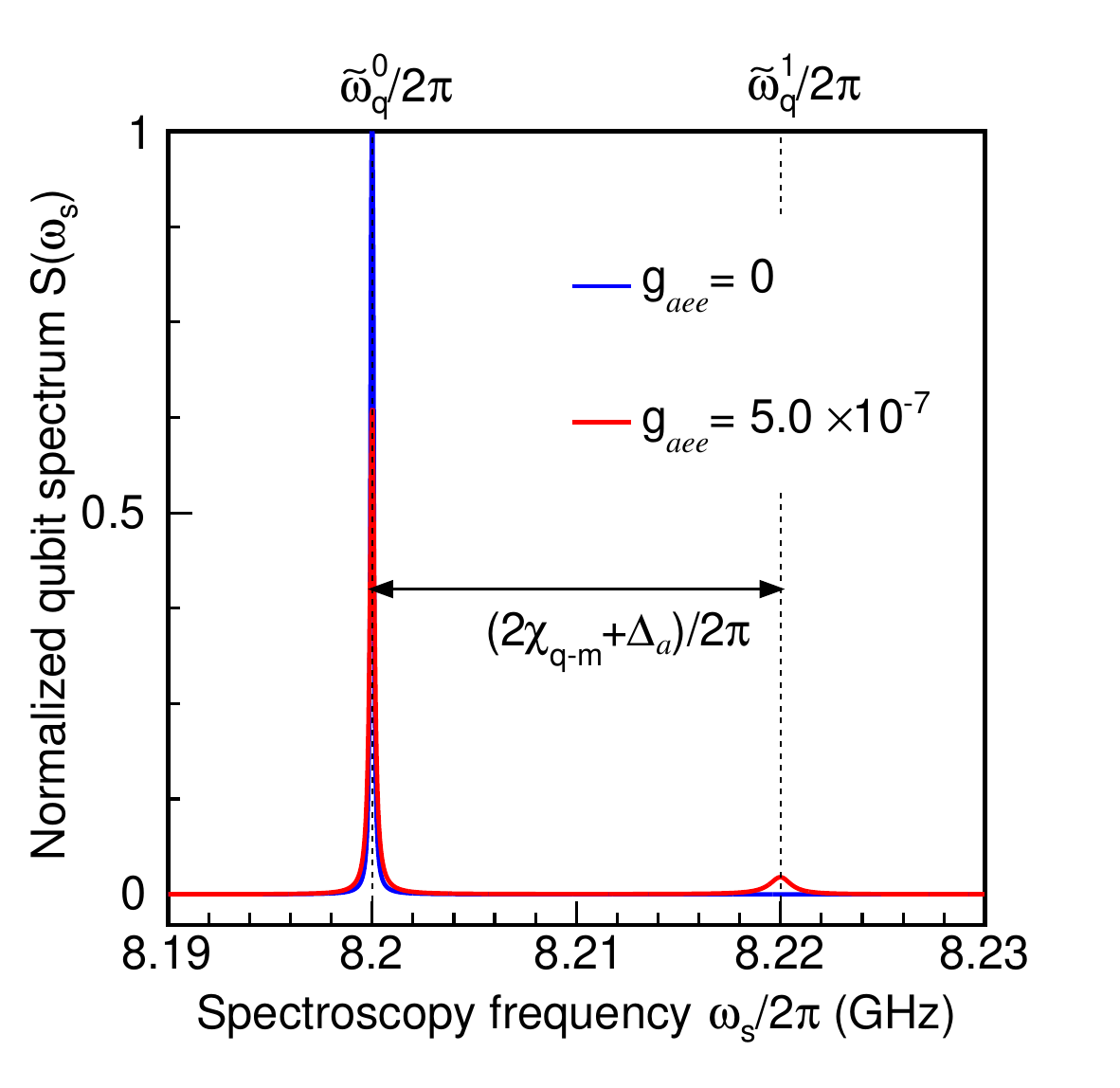}
\caption{Normalized qubit spectra $S(\omega_{\rm s})$ in the absence and presence of a coupling with axions, shown as the 
blue~($g_{aee}=0$) and red~($g_{aee}=5.0\times10^{-7}$) lines, respectively. The parameters used here are $\omega^{g}_{{\rm m}}/2\pi=8.0$~GHz, $\omega_{a}/2\pi=8.0$~GHz, $\tilde\omega_{{\rm q}}/2\pi=8.2$~GHz, $\gamma_{{\rm m}}/2\pi=1.0$~MHz, $\gamma_{{\rm q}}/2\pi=0.1$~MHz, and $\chi_{{\rm q}-{\rm m}}/2\pi=10$~MHz.}
\label{fig:expected_spec}
\end{figure}

Figure~\ref{fig:expected_spec} shows the expected qubit spectrum for the different values of the coupling constant $g_{aee}$ considering a spherical ferrimagnetic crystal of yttrium iron garnet (${\rm Y_{3}Fe_{5}O_{12}}$, YIG). 
This crystal is considered to have a diameter of $0.5$~mm with a net spin density of $\sim2.1\times10^{22}$~cm$^{-3}$ and the following realistic parameters: $\omega^{g}_{{\rm m}}/2\pi=8.0$~GHz, $\omega_{a}/2\pi=8.0$~GHz (corresponding to an axion mass $m_{a}\approx33$~$\mu$eV), $\tilde\omega_{{\rm q}}/2\pi=8.2$~GHz, $\gamma_{{\rm m}}/2\pi=1.0$~MHz, $\gamma_{{\rm q}}/2\pi=0.1$~MHz, and $\chi_{{\rm q}-{\rm m}}/2\pi=10$~MHz. 
For the calculation, the magnon number is truncated at $n_{\rm{m}}=10$, which is justified for $\overline{n}^{g(e)}_{{\rm m}}\ll1$. 
Without the axion-induced magnetic field ($g_{aee} = 0$), only one peak appears at the qubit frequency $\tilde{\omega}_{\rm q }^{(0)}=\tilde\omega_{\rm q}$. On the other hand, for $g_{aee}=5.0\times10^{-7}$, a second peak at $\tilde{\omega}_{\rm q }^{(1)}\approx\tilde{\omega}_{\rm q }^{(0)}+2\chi_{{\rm q}-{\rm m}}$ appears. This second peak corresponds to the excitation of a single magnon in the Kittel mode due to the axion DM. 
Therefore, the observation of a peak around the frequency $\tilde{\omega}_{\rm q }^{(1)}$ demonstrates the existence of the axion DM.

\section{Experiment}
In Ref.~\cite{Lachance-Quirione1603150}, owing to the strong dispersive regime between the uniform magnetostatic mode of the spherical ferrimagnetic crystal and a superconducting qubit, quanta of the collective spin excitations were observed under an additional resonant drive of the Kittel mode. 
Here, the data without an intentional drive is analyzed for the search of axion DM. More details about the experiment can be found in Ref.~\cite{Lachance-Quirione1603150}.

\subsection{Device and parameters}
As depicted in Fig.~\ref{fig:scheme}, the hybrid quantum system consists of a superconducting qubit and a single crystalline sphere of YIG, both inside a three-dimensional microwave cavity. The diameter of the YIG sphere is $0.5$~mm. 
A pair of permanent magnets and a coil are used to apply an external magnetic field $B_{0}\approx0.29$~T for the YIG sphere. The transmon-type superconducting qubit has a frequency $\tilde{\omega}_{\rm q}/2\pi\approx7.99$~GHz.

The superconducting qubit and the Kittel mode of the ferrimagnetic crystal are coherently coupled by their individual interactions with the modes of the microwave cavity~\cite{TABUCHI2016729,Tabuchi405,Lachance-Quirione1603150}. The effective coupling strength $g_{\rm{q-m}}/2\pi=7.79$~MHz is experimentally determined from the magnon-vacuum Rabi splitting of the qubit. This coupling strength is much larger than the power-broadened qubit linewidth $\gamma_{\rm{q}}$ and the magnon linewidth~$\gamma_{\rm{m}}$~\cite{Lachance-Quirione1603150}.

The frequency of the Kittel mode is set based on the current in the coil to reach the dispersive regime of the interaction between the uniform mode and the qubit. The amplitude of the detuning $\left|\Delta_{\rm{q}-\rm{m}}\right|=\left|\tilde{\omega}_{\rm{q}}-\omega_{\rm{m}}^g\right|$ is much larger than the coupling strength $g_{\rm{q}-\rm{m}}$. The qubit-magnon dispersive shift $\chi_{\rm{q-m}}/2\pi=1.5\pm0.1$~MHz and the dressed magnon frequency $\omega_{\rm{m}}^{g}/2\pi=7.94962$~GHz are obtained. 
These values are summarized in Table~\ref{tab:params}.

\begin{table*}
\caption{\label{tab:params}Parameters determined in the experiment. The error ranges indicate the 95\% confidence interval.}
\begin{ruledtabular}
\begin{tabular}{lll}
Parameter & Symbol & Value \\ \hline
Dressed magnon frequency & $\omega_{\rm{m}}^{g}/2\pi$ & 7.94962 GHz\\
ac-Stark-shifted qubit frequency & $\omega_{\rm{q}}^{(n_{\rm p}=0)}/2\pi$  & 7.99156 GHz \\
Broadened qubit linewidth & $\gamma_{\rm{q}}^{(n_{\rm p}=0)}/2\pi$ & 0.78 $\pm$ 0.03 MHz \\
Probe cavity-mode linewidth & $\kappa_{\rm{p}}/2\pi$ & 3.72 $\pm$ 0.03 MHz \\
Magnon linewidth & $\gamma_{\rm{m}}/2\pi$ & 1.3 $\pm$ 0.3 MHz \\
Qubit--probe-mode dispersive shift & $\chi_{\rm{q-p}}/2\pi$ & $-0.8 \pm 0.2$ MHz  \\
Qubit--magnon-mode dispersive shift & $\chi_{\rm{q-m}}/2\pi$ & 1.5 $\pm$ 0.1 MHz\\
Probe-mode occupancy & $\overline{n}_{\rm{p}}^g$ & 0.22 $\pm$ 0.17 \\
\end{tabular}
\end{ruledtabular}
\end{table*}

\subsection{Measurements and results}
\begin{figure}
\begin{center}
\includegraphics[width=8.5cm]{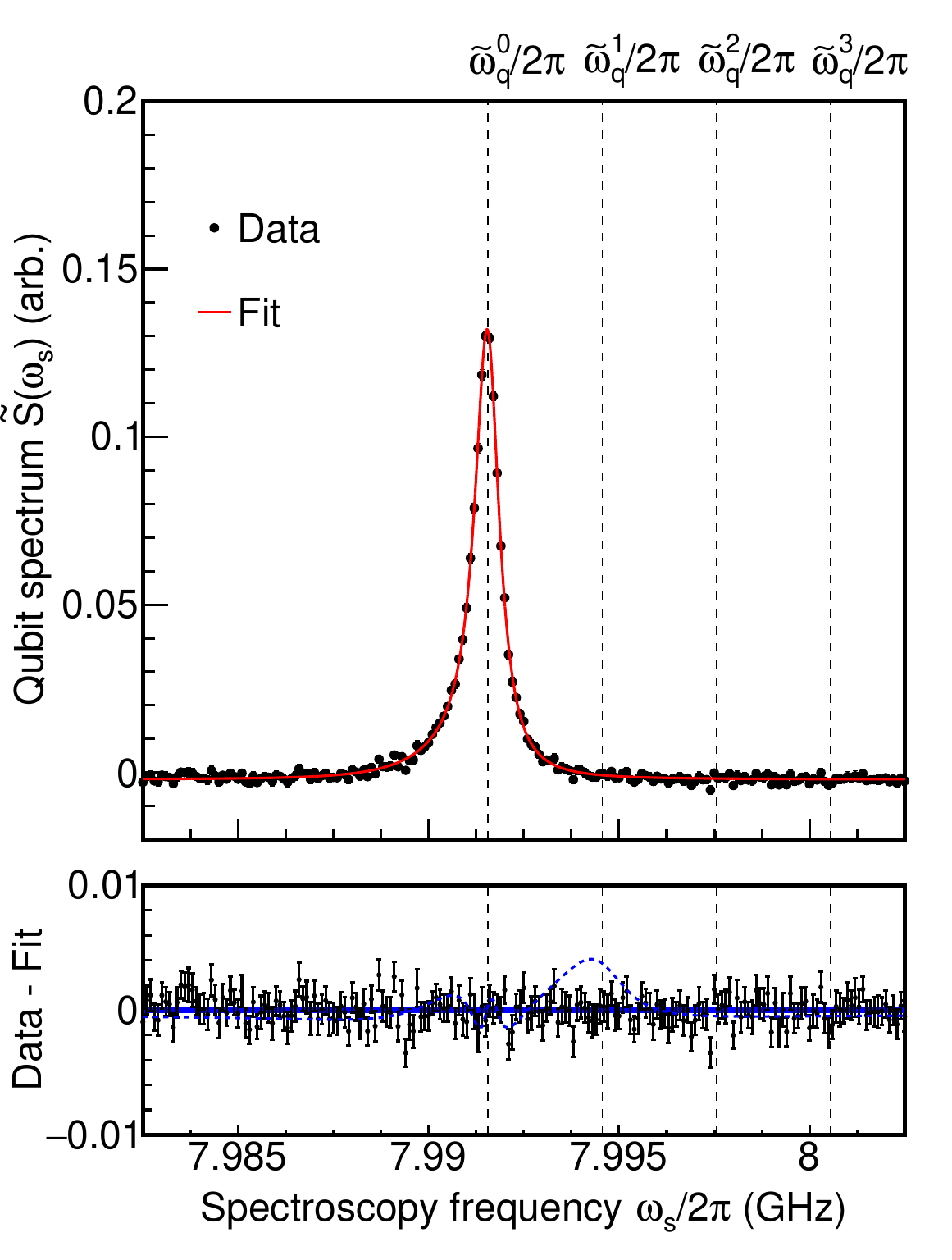}
\caption{(a)~Measured qubit spectrum $\tilde{S}(\omega_{\rm s})$ (black dots) and fit to Eq.~\eqref{eq:fit} (red line).
(b)~Residuals between the measured qubit spectrum and the fit. The blue dotted curve shows the expected residual at the 95\%-confidence-level upper limit ($\bar{n}_{\rm{m}}^{g}=0.011$) magnified by 10 times.}
\label{fig:result}
\end{center}
\end{figure}

Here we analyze the qubit spectrum obtained in the experiment of Ref.~\cite{Lachance-Quirione1603150} without an intentional excitation of the Kittel mode. The qubit absorption spectrum shown in Fig.~\ref{fig:result} was measured in the frequency range $7.9825$~GHz~$<\omega_{\rm s}/2\pi<8.0025$~GHz with a resolution of $100$~kHz. 
The measurement lasted approximately $4$~hours and each bin of $50$~measurements was averaged.

The measured spectrum is fitted according to the following equation,
\begin{equation}
\tilde{S}(\omega_{\rm s})=\mathcal{A} \sum_{n_{\rm m}=0}^{10} S_{n_{\rm m}}(\omega_{\rm s}) + S_{\rm off},
\label{eq:fit}
\end{equation}
where $\mathcal{A}$ is a scaling factor, $S_{\rm off}$ is an offset, and $S_{n_{\rm{m}}}(\omega_{\rm s})$ is the contribution of the qubit spectrum with $n_\mathrm{m}$ magnons excited in the uniform magnetostatic mode. 
The small asymmetry of the qubit spectrum corresponding to $n_\mathrm{m}=0$ is due to the finite photon occupancy of the microwave cavity mode that is used to probe the qubit. 
The effect of the photon occupation in the probe mode can be considered as follows:
\begin{equation}
S_{n_{\rm m}}(\omega_{\rm s}) \thickapprox S_{n_{\rm m}, n_{\rm p}=0}(\omega_{\rm s}) + \mathcal{B} \times S_{n_{\rm m}, n_{\rm p}=1}(\omega_{\rm s}),
\end{equation}
where $\mathcal{B}=0.03$ is the relative spectral weight between the one-photon and the zero-photon peaks. To consider the ac Stark shift of the qubit frequency by the photons in the probe mode used to measure the qubit spectrum, the following is substituted as 
\begin{equation}
\tilde{\omega}_{\rm q} \rightarrow  \omega_{\rm q}^{(n_{\rm p}=0)}=\tilde{\omega}_{\rm q}+B_{\rm p},
\end{equation}
where $\omega_{\rm q}^{(n_{\rm p}=0)}$ is the ac-Stark-shifted qubit frequency with the Kittel mode in the vacuum state. The qubit linewidth with the Kittel mode in the vacuum state is substituted to
\begin{equation}
\gamma_{\rm q} \rightarrow \gamma_{\rm q}^{(n_{\rm p}=0)}=\gamma_{\rm q}+\kappa_{\rm p}D_{\rm p}^{\rm ss},
\end{equation}
where $\gamma_{\rm q}^{(n_{\rm p}=0)}$ is the linewidth increased by the measurement-induced dephasing from photons in the probe mode.
The parameters fixed in the fit of the qubit spectrum are the qubit frequency $\tilde{\omega}_{\rm{q}}$, the qubit linewidth~$\gamma_{\rm{q}}$, the probe mode occupancy $\overline{n}_{\rm{p}}^g$, the qubit--probe-mode dispersive shift $\chi_{\rm{q-p}}$, the probe cavity-mode linewidth~$\kappa_{\rm{p}}$, the qubit--magnon-mode dispersive shift~$\chi_{\rm{q-m}}$, the magnon linewidth~$\gamma_{\rm{m}}$, and the drive detuning $\Delta_{a}$.

To consider the uncertainties of the experimentally measured parameters, the chi-square function $\chi^{2}$ was defined with nuisance parameters $\alpha_{j}$ as
\begin{align}
\chi^{2}&\equiv\sum_{i=0}^{n}\frac{\tilde{S}_{i}-\tilde{S}\left(\omega_{\rm s}^{(i)},\gamma_{\rm{q}}',{\overline{n}_{ \rm{p} }^g}',\chi_{\rm{q-p}}', {\kappa_{\rm{p}}}', \chi_{\rm{q-m}}', \gamma_{\rm{m}}'\right)}{\sigma_{\tilde{S}_{i}}} \notag \\
 &+\sum_{j=0}^{5}\alpha_{j}.
\end{align}
with
\begin{align*}
\gamma_{\rm{q}}'&=\gamma_{\rm{q}}-\alpha_{0}\sigma_{\gamma_{\rm{q}}},\\
{\overline{n}_{ \rm{p} }^g}' &=\overline{n}_{\rm{p} }^g-  \alpha_{1}\sigma_{ \overline{n}_{\rm{p}}^g },\\
\chi_{\rm{q-p}}' &=\chi_{\rm{q-p}}- \alpha_{2}\sigma_{\chi_{\rm{q-p}}},\\
{\kappa_{\rm{p}}}' &=\kappa_{\rm{p}}- \alpha_{3}\sigma_{\kappa_{\rm{p}}},\\
\chi_{\rm{q-m}}'&=\chi_{\rm{q-m}}-\alpha_{4}\sigma_{\chi_{\rm{q-m}}},\\
\gamma_{\rm{m}}'&=\gamma_{\rm{m}}-\alpha_{5}\sigma_{\gamma_{\rm{m}}},
\end{align*}
where $\tilde{S}_{i}$ and $\sigma_{\tilde{S}_{i}}$ are the average and standard deviation of the experimentally measured spectrum for bin $i$ for a spectroscopy frequency $\omega_{\rm s}^{(i)}$, respectively. Systematic errors of $\gamma_{\rm{q}}$, $\overline{n}_{\rm{p} }^g$, $\chi_{\rm{q-p}}$, $\kappa_{\rm{p}}$,  $\chi_{\rm{q-m}}$, and $\gamma_{\rm{m}}$ are given by $\sigma_{\gamma_{\rm{q} }}$, $\sigma_{ \overline{n}_{\rm{p}}^g }$, $\sigma_{\chi_{\rm{q-p}}}$, $\sigma_{\kappa_{\rm{p}}}$, $\sigma_{\chi_{\rm{q-m}}}$, and $\sigma_{\gamma_{\rm{m}}}$, respectively.

The results of the fitted data are presented in Fig.\,\ref{fig:result}, where the average number of magnons is fixed to $\overline{n}_{\rm{m}}^g=0$, which is equivalent to $g_{\rm{eff}}=0$. From this, $\chi^{2}$ was reduced to 167.1/193. This is consistent with the null hypothesis and there was
no significant excess in the residuals found for the frequency $\tilde{\omega}_{\rm q}^{(1)}=\tilde{\omega}_{\rm q}^{(0)}+2\chi_{\rm{q}-\rm{m}}$. 
As a result, a 95\% confidence level is set for the upper limit of $g_{aee}$.
The upper limit of the average number of magnons $\overline{n}_{{\rm limit}}$ was calculated as follows:
\begin{align}
\frac{\int_{0}^{\overline{n}_{{\rm limit}}}L\ {\rm d}\overline{n}_{\rm{m}}^g}{\int_{0}^{\infty}L\ {\rm d}\overline{n}_{\rm{m}}^g}=0.95,\label{xitest}
\end{align}
where $L$ is defined as follows:
\begin{equation}
L\equiv\exp\left(-\frac{\chi^{2}(\overline{n}_{\rm{m}}^g)-\chi_{\rm{min}}^{2}}{2}\right).
\end{equation}
The chi-square function $\chi^{2}(\overline{n}_{\rm{m}}^g)$ is calculated by varying $\overline{n}_{\rm{m}}^g$, while $\chi_{\rm min}^{2}$ is the minimum $\chi^{2}$. From this, $\overline{n}_{{\rm limit}}=1.1\times10^{-2}$ is obtained. 
The expected residuals calculated with $\overline{n}_{{\rm limit}}$ is shown in Fig.~\ref{fig:result}.

\begin{figure}[]
\centering
\includegraphics[width=8.5cm]{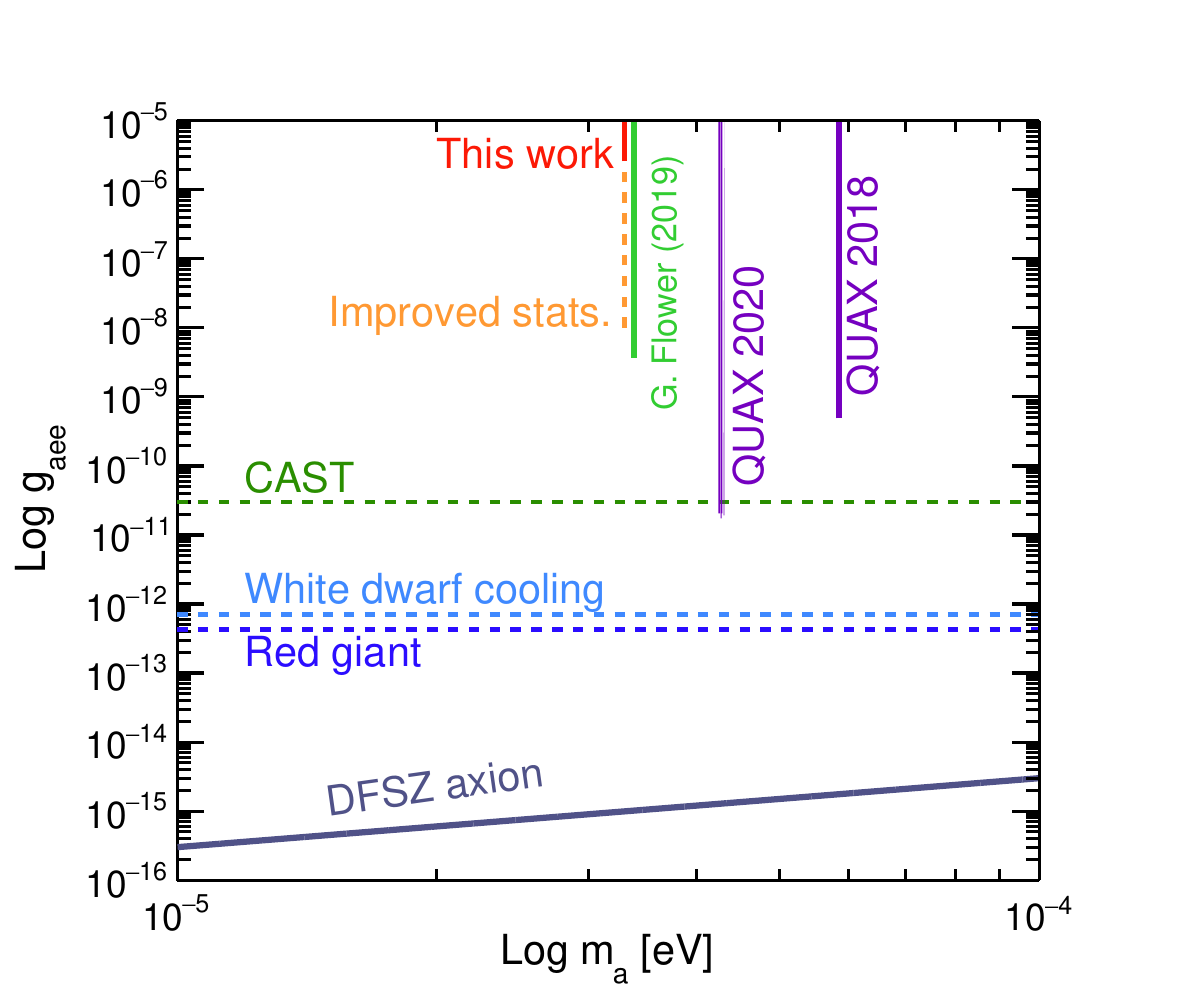}
\caption{Constraints on the coupling constant $g_{aee}$ between axions and electrons. The region excluded by this work with 95\% confidence is shown in red. The mass range is $33.117$~$\mu$eV~$<m_{a}<33.130$~$\mu$eV. The purple and green lines are the upper limit from QUAX~\cite{QUAX,crescini2020axion} and G.Flower et\,al~\cite{Tober}. Indirect astrophysical bounds from the solar axion search (CAST experiment~\cite{Barth:2013sma}), white dwarf cooling~\cite{Corsico}, and red giants~\cite{PhysRevLett.111.231301} are illustrated with dashed lines. The expected coupling constant for the DFSZ model is represented by a solid line. The orange dashed line shows the predictions for the future work. 
}
\label{fig:limit}
\end{figure}

The amplitude of the effective magnetic field is given by $B_a\sin\theta$, where $\theta$ is the angle between the direction of the external magnetic field $\bm{B_0}$ and the direction of the axion-induced effective magnetic field. 
From Eqs.~\eqref{eq:geff_Ba} and \eqref{eq:geff_and_n}, the 95\%-confidence-level upper limit on the amplitude of the effective magnetic field at $m_{a}=33.123$~$\mu$eV can be determined as follows:
\begin{equation}
B_{a}\,{\rm sin}\theta<8.2\times10^{-15}\ {\rm T}.
\end{equation}
\ikeda{We took the distribution of the axion-induced effective magnetic field described in Ref.~\cite{Barbieri:2016vwg,QUAX, Tober, crescini2020axion}.
The minimum 
$\sin\theta$ value during 4 hours operation is 0.097. From Eq.~\ref{axmag}, the upper limit of the axion-electron coupling constant is obtained as
\begin{equation}
g_{aee}<2.6\times10^{-6} \ ,
\end{equation}
using the conventional galactic density of DM $\rho_{{\rm DM}}=0.45$~GeV/cm$^{3}$~\cite{0954-3899-41-6-063101} and $v_{\rm tot}=220$~km/sec\footnote{\ikeda{More precisely, the solar system is moving with a velocity $|\bm{v}_{{\rm solar}}|\sim220$~km/s in the Galaxy~\cite{2012ApJ...759..131B} and we observe the relative velocity of the axion DM as $\bm{v}_{{\rm tot}}=\bm{v}_{{\rm solar}}+\bm{v}$. This is further discussed in Appendix~\ref{apn}.}}. The similar spectrum fitting was conducted to the range of $33.117$~$\mu$eV~$<m_{a}<33.130$~$\mu$eV in the same external magnetic field.
The constraint is plotted in Fig.~\ref{fig:limit} and is compared with other previously established bounds on the axion--electron coupling constant.}




\section{Discussion}
Although this work set the best upper limits for the axion mass $33.117$~$\mu$eV$<m_{a}<33.130$~$\mu$eV with the direct axion--electron interaction search, several orders of magnitude improvements are required to reach the theoretical predictions.
We discuss some possibilities to improve the sensitivity here.
\ikeda{In our system, the magnon excited state is detected due to the thermal photons at 10~mK, corresponding to 1.7$\times$10$^{-3}$~aW. While the thermal noise limits the sensitivity finally, our experiment did not reach such sensitivity due to the broadened magnon line width. Therefore, the most straightforward improvement is to increase the statistics with a longer measurement time.}
The data used for this axion search was originally taken for a different purpose \cite{Lachance-Quirione1603150}. The data acquisition time 
was roughly 4 hours for the spectroscopy window $[7.9825-8.0025~\mathrm{GHz}]$ with a resolution of $100$ kHz. 
Assuming one week data taking limiting the frequency window to $[7.9860-7.9960~\mathrm{GHz}]$, a relevant window for the measurement of the single-magnon excited state, a 100-hold statistics increase can easily be made.
In addition, data acquisition over several days could help to uncover the expected daily modulation for the axion signal.

Another 100 times statistics increase for a given measurement time can be made by further limiting the the spectroscopy window to a few bins which correspond to the magnon linewidth resolution. It should be noted that this approach requires some improvements on the qubit-magnon coupling condition to narrow the relevant linewidth.
With these improvements, it 
is expected to lower the bound to a coupling strength \ikeda{$g_{aee}\sim 10^{-8}$}. 

The sensitivity can also be improved by 
increasing the number of electron spin targets. Coupling $N$ pieces of YIG spheres in the uniform mode to the superconducting qubit increases the effective coupling constant by a factor of $\sqrt{N}$. The QUAX experiment deployed ten pieces of YIG spheres of 2.1~mm diameter and succeeded in increasing the number of electron spin targets~\cite{QUAX}. This technique can also be used for our case and would be expected to improve the sensitivity for the coupling strength.

As aforementioned, Refs.{~\cite{QUAX, Tober, crescini2020axion}} showed the upper bound of the axion-electron coupling using the magnon in the spherical ferrimagnetic crystals. Even though this method directly measures the magnon number, the emitted electromagnetic radiation is measured from a microwave cavity where one cavity mode is hybridized with one or multiple uniform magnetostatic modes of ferrimagnetic spheres. Therefore, the background of magnon detection is different. 

\section{Conclusion} 
Magnons can be utilized for exploring the axion 
DM~\cite{QUAX, Tober, crescini2020axion} and 
gravitational waves~\cite{Ito:2019wcb,Ito:2020wxi}.
In particular, the QND detection of magnons was achieved using a hybrid
quantum system consisting of a superconducting
qubit and a spherical ferrimagnetic crystal~\cite{Lachance-Quirione1603150}.
We applied to the direct axion search based on the
axion--electron coupling and analyzed the background data.
No significant signal was detected, and an upper limit of the 95\% confidence level was set to be 
\ikeda{$g_{aee}<2.6\times10^{-6}$} for the axion--electron coupling coefficient for the axion mass $33.117$~$\mu$eV$<m_{a}<33.130$~$\mu$eV. 
The sensitivity is presently limited by statistics. 
Increasing the dispersive shift or reducing the power-broadened qubit linewidth and the magnon linewidth will
lower the upper bound on the axion--electron coupling.

\section*{Acknowledgement}
We would like to thank Yasunobu Nakamura and Dany Lachance-Quirion for providing the data used in this paper.
We are also grateful to them for valuable comments.
We acknowledge N. Crescini and C. C. Speake for their useful discussions.
A.\,I.\, was supported by National Center for Theoretical Sciences.
J.S. was supported by JSPS KAKENHI Grant Numbers JP17H02894, JP17K18778, and JP20H01902. K.\,M.\, was  supported by JSPS KAKENHI Grant Numbers 26104005, 16H02189, 19H05806 
Y. S. was partially supported by Gunma university for the promotion of scientific research, JSPS KAKENHI (Grant Nos.  19K14636 and 21H05599) and JST PRESTO (Grant No. JPMJPR20M4).

\appendix
\section{\itou{Relative velocity of axion DM to the Earth}}
\label{apn}
\itou{Although we used the model of the effective magnetic field induced by the axion DM presented 
in~\cite{Barbieri:2016vwg,QUAX, Tober, crescini2020axion} 
in the main body, here we give more precise discussion 
about it.}
The relative velocity of the axion DM to the Earth can be written as
\begin{align}
\bm{v}_{{\rm tot}}=&\,v_{{\rm solar}}(\sin\alpha,0,\cos\alpha)\nonumber \\
&+v(\cos\xi,\sin\xi\cos\phi,\sin\xi\sin\phi),  \label{vtot}
\end{align}
where $v_{{\rm solar}}=220$~km/s~\cite{2012ApJ...759..131B} is the velocity of the solar system in the Galaxy and $\alpha$ denotes the angle between $\bm{v}_{{\rm solar}}$ and the external magnetic field $B_{0}$. 
\itou{The second term represents random motion of the 
axion DM.}
We assumed that the virial velocity in the Galaxy $\bm{v}$ has an isotropic Gaussian distribution with a dispersion $\sigma=270$~km/s.
Then one can not predict the direction of the relative velocity of the axion DM in advance. Therefore we use the expectation value by taking average over the angles.%
\footnote{In~\cite{Barbieri:2016vwg,QUAX, Tober, crescini2020axion}, it is assumed that 
the direction of the relative velocity of the axion DM to the Earth 
is only given by the first term in Eq.\,(\ref{vtot}) by dropping the second term.}
The expectation value of $v_{{\rm tot}}\sin\theta$ can be evaluated, where $\theta$ is the angle between $\bm{v}_{{\rm tot}}$ and $B_{0}$, as
\begin{widetext}
\begin{align}
\langle v_{{\rm tot}}\sin\theta\rangle=\frac{1}{4\pi}\int_{0}^{2\pi}d\phi\int_{-1}^{1}d(\cos\xi)\ \frac{4\pi}{(2\pi\sigma^{2})^{3/2}}\int_{0}^{\infty}dv\ v^{2}e^{-\frac{v^{2}}{2\sigma^{2}}}\sqrt{\left(v_{{\rm solar}}\sin\alpha+v\cos\xi \right)^{2}+v^{2}\sin\xi^{2}\cos\phi^{2}}.
\end{align}
\end{widetext}
For the most conservative case $\alpha=0$, this results in $\langle v_{{\rm tot}}\sin\theta\rangle=338$~km/s. 
\ikeda{In this case, our upper limit of the axion--electron coupling constant is modified to $g_{aee}<1.6\times 10^{-7}$.}

\bibliographystyle{apsrev4-2}
\bibliography{main}

\end{document}